# Ni-Mn-Ga films in the austenite and the martensite structures at room temperature: Uniaxial texturation and epitaxial growth

Jérémy Tillier[1], Antinéa Einig[2], Daniel Bourgault[1], Philippe Odier[1], Luc Ortega[1], Sébastien Pairis[1], Laureline Porcar[1], Paul Chaumeton[1], Nathalie Caillault[2], Laurent Carbone[2]
[1]Institut Néel/CRETA, CNRS, 25 avenue des martyrs, BP166, 38042 Grenoble Cedex 9, France
[2]Schneider Electric France, 37 quai Paul Louis Merlin, 38050 Grenoble Cedex 9, France

**ABSTRACT**

Ni-Mn-Ga films in the austenite and the martensite structures at room temperature have been obtained using the DC magnetron sputtering technique. Two elaboration processes were studied. A first batch of samples was deposited using a resist sacrificial layer in order to release the film from the substrate before vacuum annealing. This process leads to polycrystalline films with a strong (022) fiber texture. The martensitic phase transformation of such polycrystalline freestanding films has been studied by optical and scanning electron microscopy. A second batch of samples was grown epitaxially on (100)MgO substrates using different deposition temperatures. The texture has been analyzed with four-circle X-ray diffraction. Epitaxial films crystallized both in the austenite and the martensite structures at room temperature have been studied.

**INTRODUCTION**

The ternary intermetallic compound $Ni_2MnGa$ has gained considerable attention, as it exhibits an interesting combination of thermoelastic and magnetic properties. In addition to the conventional shape memory effect, this system shows a magnetic field induced rearrangement of martensitic twinned structures commonly called as magnetic shape memory (MSM) effect [1,2]. Recent investigations in $Ni_2MnGa$ bulk single crystals revealed considerable shape changes up to 10% and an interesting response time below 1 ms [3-6]. So Ni-Mn-Ga alloys have a great potential of application as microactuators or microsensors for micro-electro-mechanical systems. However, the bulk alloy is too brittle to be machined in a required shape and Ni-Mn-Ga alloys have to be prepared directly in film shape [7]. So far, Ni-Mn-Ga films have been grown by means of various techniques such as molecular beam epitaxy [8], sputtering [9-11] and pulsed laser deposition [12]. The texturation of the film is a key parameter to maximize intensity of the MSM effect. Recent studies on epitaxially grown Ni-Mn-Ga films showing a biaxially textured 7M martensite structure exhibit MSM effect at room temperature [13].
In the first part of the discussion, the texture of a freestanding Ni-Mn-Ga film austenitic at room temperature is analyzed. The martensitic phase transformation is characterized both by optical and scanning electron microscopy. The second part focuses on the epitaxial growth of Ni-Mn-Ga austenite at high temperature. The crystallographic structure and texture of a martensitic film at room temperature is also discussed.

**EXPERIMENT**

The freestanding films have been deposited on photoresist (Shipley S1818) sacrificial layers, which were spin-coated on polycrystalline alumina substrates. The Ni-Mn-Ga films with a thickness of about 5μm were synthesized using the direct current (DC) magnetron sputtering technique with a 2-in. target of Ni-Mn-Ga alloy. Details of deposition parameters are described in reference [14]. After deposition, the Ni-Mn-Ga/S1818/Al$_2$O$_3$ composites were placed into an acetone bath to strip the resist sacrificial layer. Then, the freestanding films fastened by alumina ceramic pieces were annealed at 1093 K for 6 h under a vacuum below 10$^{-3}$ Pa followed by furnace cooling to achieve homogenization and ordering.

The epitaxially grown films have been deposited on heated (100)MgO monocrystals with a substrate rotation of 5 rpm. The Ni-Mn-Ga films with a thickness of about 500 nm were sputtered from a 2-in. target of Ni-Mn-Ga alloy using a DC magnetron confocal cathode located at 10.5 cm from the substrate. The working pressure was kept at 0.8 Pa by controlling an Ar gas flow of 25 sccm. The sputtering power was optimized to obtain a deposition speed of 1μm.h$^{-1}$.

The crystallographic structures of the films were analyzed using an X-ray diffraction
(XRD) instrument (Siemens D5000) with Cu Kα radiation. The microstructures were examined by scanning electron microscopy (SEM JEOL 8480A). Indicated film compositions have been determined by EDX. Optical observations were realized using a Zeiss microscope. The film was cooled down with a hand-made helium circulation cryostat [15]. The film textures have been studied using a four-circle XRD instrument (Seifert MZ IV) using a parallel Cu Kα beam and a rear monochromator.

**DISCUSSION**

**Freestanding films with (022) fiber texture**

The selected annealing parameters allow a complete crystallization of the films. After vacuum heat-treatment at 1093K for 6h, the XRD scan of the Ni$_{52.5}$Mn$_{24}$Ga$_{23.5}$ annealed film shown in figure 1.a. can be indexed by the Heusler type Fm-3m cubic structure with a lattice parameter of 5.797 Å. The film is thus crystallized in the austenitic structure at room temperature.

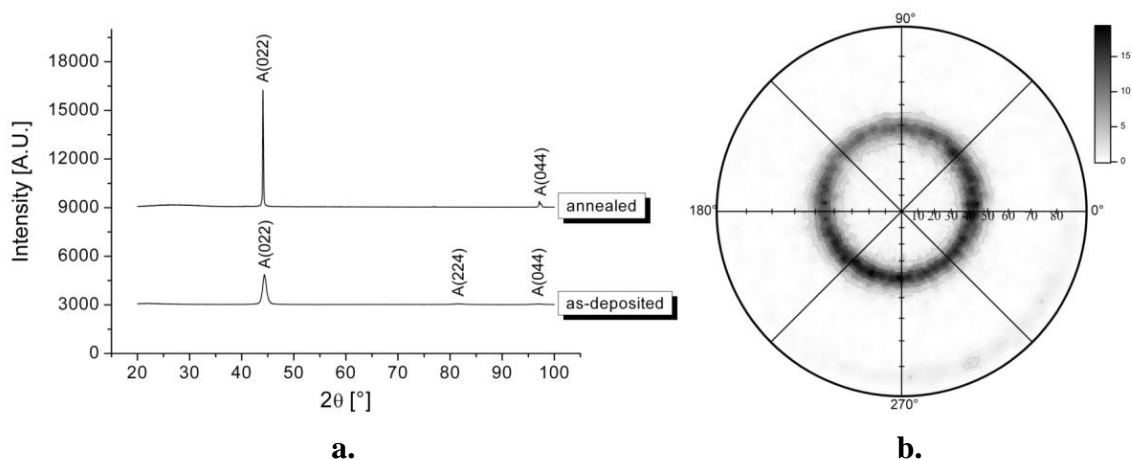

                a.                 b.

**Figure 1.a.** θ-2θ XRD scans of an as-deposited and an annealed Ni$_{52.5}$Mn$_{24}$Ga$_{23.5}$ film. The indices indicate the peak positions of the austenite. **b.** (004) pole figure of the annealed film.

In this cubic phase only the (022) and (044) peaks were found. The film demonstrates a strong out-of-plane texture along the [022] direction. The XRD pattern did not show additional peaks, hence no precipitation of others phases occurs. The (004) pole figure of the freestanding film is shown in figure 1.b. The deviation of the pole figure from the centrosymmetrical position is due to imprecise positioning of the freestanding film on the substrate. The ring at the tilt angle ψ=45° demonstrates the absence of in-plane texture. The freestanding film exhibits a strong (022) fiber texture.

The martensitic phase transformation of such freestanding Ni-Mn-Ga films showing (022) fiber texture has been studied by temperature-dependent optical microscopy. Within the austenitic structure, the film surface is flat and shinny (figure 2.a and 2.f.). When the film is cooled down through the martensitic transition (figure 2.b.), some areas with a different roughness appear. The observed temperature-dependent appearance of rough areas is ascribed to martensitic grains originating from different nucleation sites which grow at the expense of the austenite. By heating the film from the martensite structure (figure 2.c. and 2.d.), austenite grains grow in the martensite (figure 2.e.) and the film shows again its flat surface with shinny appearance (figure 2.f.). Figure 2.b and 2.e show different nucleus growing at the expense of the other phase. It agrees with a first order phase transition [10].

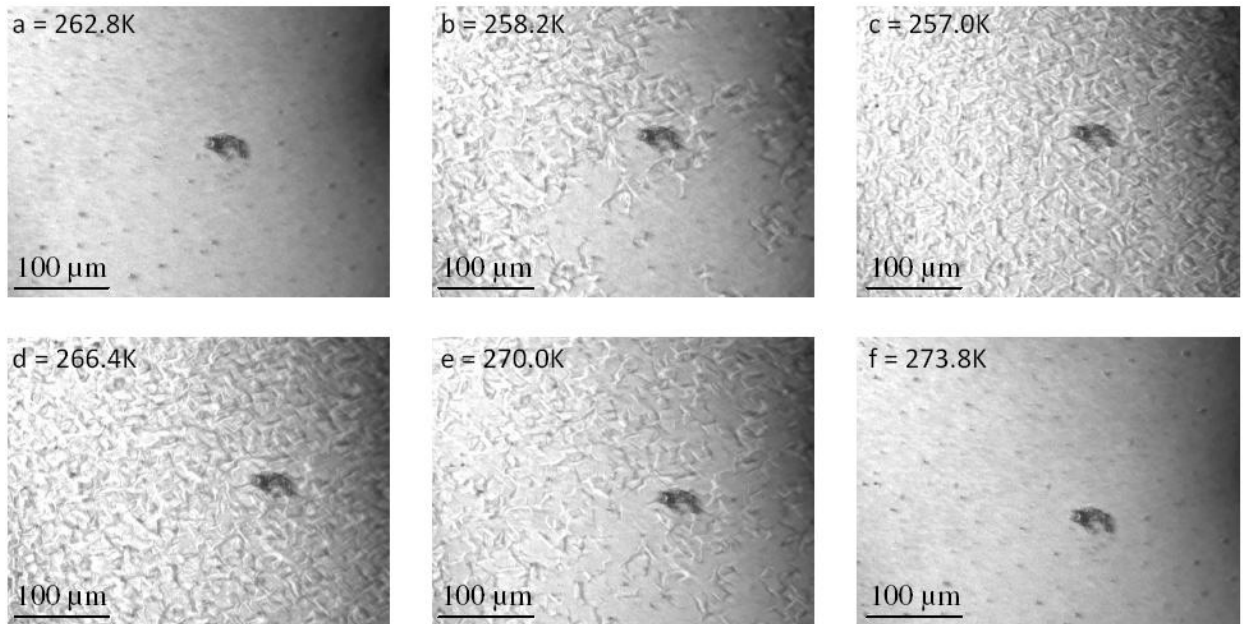

**Figure 2.** Optical observations of the temperature-dependent surface morphology for a $Ni_{52.5}Mn_{24}Ga_{23.5}$ film in the austenite structure at RT.

The roughness appearance comes from the twinning of the martensite. In order to form a habit plane, the martensitic transformation requires a tilt and rotation and leads to crinkle the film at some grain boundaries and especially in the grains. It can be seen on figure 3, which shows the temperature-dependent SEM surface morphology of the $Ni_{52.5}Mn_{24}Ga_{23.5}$ freestanding film. Such fiber textured freestanding Ni-Mn-Ga films demonstrate a fully reversible shape memory surface. Due to a sharp increase of roughness in the martensite, a strong reflective index jump is expected. The fiber textured Ni-Mn-Ga freestanding films might be good candidates for micro-opto-electro-mechanical systems [14].

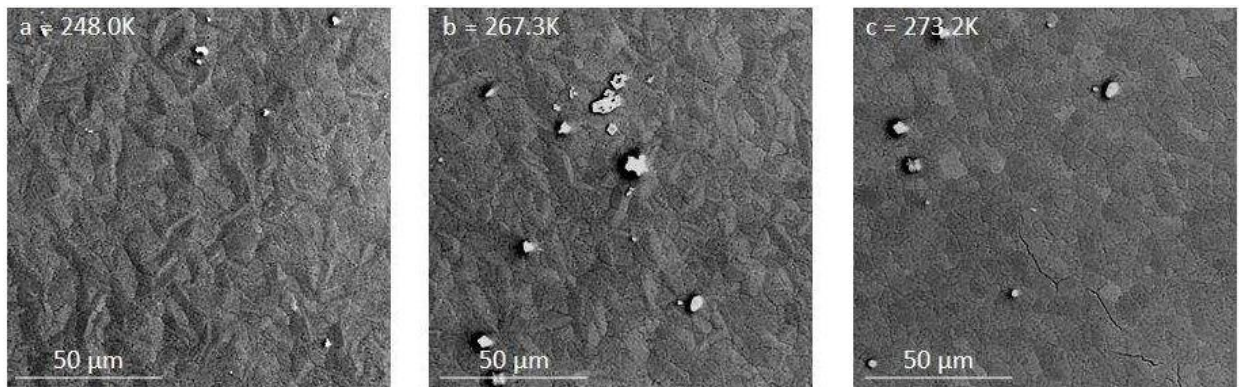

**Figure 3.** Typical secondary electron SEM images of the surface morphology for the martensite structure (a), a mix of martensite and austenite (b) and the austenite structure (c).

**Epitaxially grown films**

The θ-2θ XRD patterns of a batch of samples deposited at different temperatures is shown in figure 4.a. All the films are crystallized in the cubic Fm-3m austenite structure. Two peaks in the measured patterns originate from the single crystalline (100)MgO substrate. A XRD scan of the (100)MgO substrate is depicted in figure 4.a. From the cell parameters of the cubic substrate and Ni-Mn-Ga austenite, a Ni-Mn-Ga(001)[100]//MgO(100)[011] epitaxial orientation is expected [10,13]. At 300°C, the substrate temperature is too low to allow the epitaxial growth. In fact, no (00l) reflections of the austenite are observed. A partial epitaxy is obtained at 410°C and 460°C.

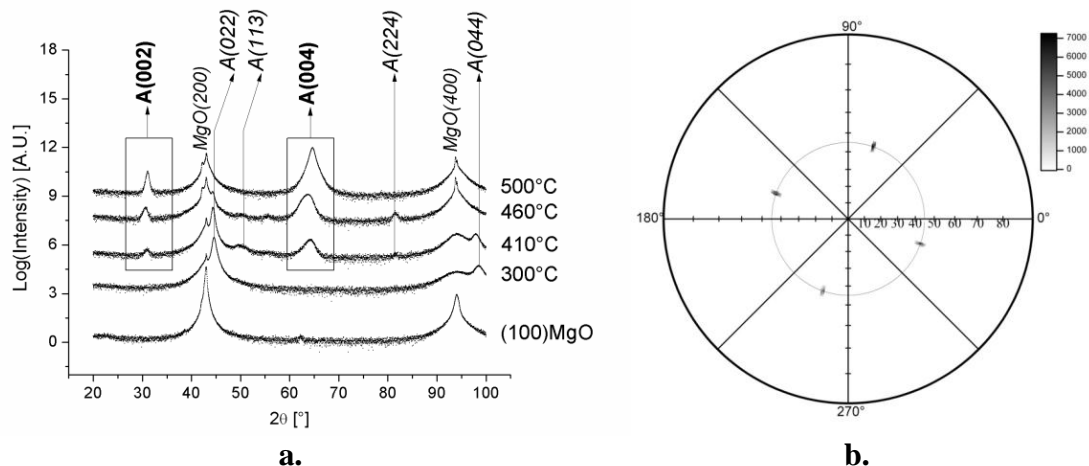

**Figure 4.a.** θ-2θ XRD scans of Ni-Mn-Ga films deposited at different temperatures. The indices indicate the peak positions for polycrystalline austenite and monocrystalline (100)MgO. The θ and 2θ angles were dissociated [(θ-1)-2θ] to reduced intensity of the (100)MgO substrate reflections. **b.** (022) pole figure of the austenitic film deposited at 500°C.

For a substrate temperature of 500°C, only the (002) and (004) peaks of the austenite are found. It indicates a strong out-of-plane texture along the [001] direction of the austenitic film. The (022) pole figure of this cubic film is shown on figure 4.b. The four strong peaks at the tilt angle ψ=45° spaced by 90° for the rotation angle Φ demonstrate the biaxial texture of the cubic austenite. For the selected sputtering parameters, epitaxial growth of the cubic austenite on

(100)MgO occurs at 500°C. The full width at half maximum (FWHM), which represents the average in-plane disorientations for the rotation angle Φ and the average out-of-plane disorientations for the tilt angle ω have been determined to be 0.60° and 0.55°, respectively (not shown). It demonstrates the good crystallographic agreement between the austenite and the (100)MgO substrate.

An epitaxial film exhibiting a martensite structure at RT has been deposited using another Ni-Mn-Ga ternary target to achieve the $Ni_{59}Mn_{22}Ga_{19}$ composition. The film with a thickness of around 500nm was deposited on a (100)MgO substrate at 500°C. In the Bragg-Brentano geometry (θ-2θ) and except the (200) and (400) reflections of the (100)MgO substrate, only two strong peaks were found at 2θ=26.55° and 54.67° (figure 5.a.). No (00l) peaks of the austenite are observed in the XRD scan. The epitaxial austenite structure crystallized during deposition at 500°C is transformed into martensite. The two strong peaks observed at RT correspond to the (002) and (004) reflections of a non-modulated (NM) martensite structure. In fact, the maximal intensity for the martensite (022) peak has been found for 2θ=42.86° at the tilt angle ψ=51° (not shown). It agrees with a NM tetragonal martensite structure described by a short cell parameter a=5.42Å and a long axis c=6.71Å. Figure 5.b. shows the (022) pole figure of the NM martensite. The four peaks at the tilt angle ψ=51° spaced by 90° for the rotation angle Φ can not originate from the MgO substrate and demonstrate the biaxial texture of the NM tetragonal martensite. A very intense and sharp peak can be seen at the center of the Wulff projection in figure 5.b. This reflection is attributed to the (200) planes of the (100)MgO substrate which diffract at a 2θ angle of 42.88°, very close to the 2θ angle of 42.86° use to realize the (022) pole figure of the NM martensite.

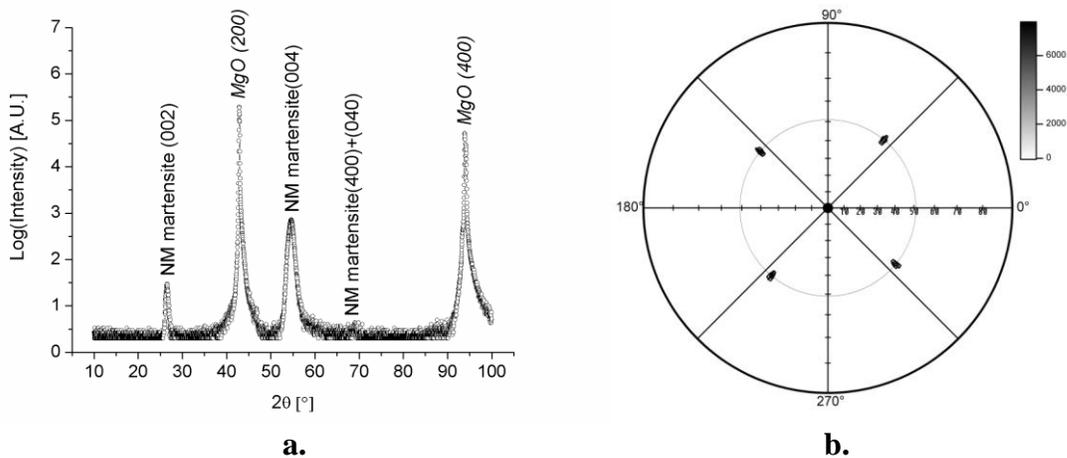

**a.**   **b.**

**Figure 5.a.** θ-2θ XRD scan of the $Ni_{59}Mn_{22}Ga_{19}$ epitaxial film **b.** (022) pole figure of the $Ni_{59}Mn_{22}Ga_{19}$ epitaxial film showing a non-modulated martensite structure at room temperature.

Whatever the temperature and the direction of the applied magnetic field, no evidence of MSM effect is observed in this film (not shown). It is imputable to the non-modulated structure of the martensite. In this structure, the twinning stress is known to exceed the magnetic induced stress and, consequently, the MSM effect is not observed in it [16]. Obtaining the NM martensite structure at RT is ascribed to the film composition which leads to a high valence electron density e/a of 8. Our group is now working on a PVD co-sputtering process to precisely master the film composition and obtain biaxially textured modulated martensites at room temperature.

## CONCLUSIONS

Freestanding Ni-Mn-Ga films with strong (022) fiber texture have been successfully prepared. Due to the polycrystalline nature and the absence of in-plane texture, these films present no interest for the MSM effect. Nevertheless, they demonstrate a fully reversible shape memory surface showing two different roughness states. It can be a good candidate for micro-opto-electro- mechanical systems. For the selected deposition speed of $1\mu m.h^{-1}$, the epitaxial growth of the austenite occurs for a (100)MgO substrate temperature of 500°C.

## ACKNOWLEDGMENTS

The NANOsciences and Matière Condensées, Matériaux et Fonctions (MCMF) departments of the Institut Néel (CNRS, Grenoble, France) are gratefully acknowledge for their contribution in this work.